\documentclass[9pt,twocolumn,twoside]{osajnl}

\usepackage[]{graphicx}
\usepackage{amssymb}
\usepackage{epstopdf}
\usepackage{xcolor}
\usepackage{amsmath}
\usepackage{gensymb}
\usepackage{tabularx}
\usepackage{color}

\def\micron{$\mu$m}

\def\mathrelfun#1#2{\lower3.6pt\vbox{\baselineskip0pt\lineskip.9pt
  \ialign{$\mathsurround=0pt#1\hfil##\hfil$\crcr#2\crcr\sim\crcr}}}
\def\simlt{\mathrel{\mathpalette\mathrelfun <}}

\journal{ao} 

\setboolean{shortarticle}{false} 

\title{Millimeter-Wave Broadband Anti-Reflection Coatings Using Laser Ablation of Sub-Wavelength Structures}

\author[1]{Tomotake Matsumura}
\author[2]{Karl Young}
\author[2]{Qi Wen}
\author[2]{Shaul Hanany}
\author[3]{Hirokazu Ishino}
\author[4]{Yuki Inoue}
\author[5,6]{Masashi Hazumi}
\author[7]{J\"{u}rgen Koch}
\author[7]{Oliver Suttman}
\author[7]{Viktor Sch\"{u}tz}

\affil[1]{Japan Aerospace Exploration Agency (JAXA) - Institute of Space and Astronautical Science (ISAS), Rm1624, 3-1-1 Yoshinodai, Chuo, Sagamihara, Kanagawa 252-5210, Japan}
\affil[2]{School of Physics and Astronomy, and Minnesota Institute for Astrophysics, University of  Minnesota/Twin Cities, 116 Church St. SE Minneapolis, MN 55455, USA}
\affil[3]{Okayama University, 3-1-1 Tsushima-naka, Kita-ku, Okayama 700-8530 Japan}
\affil[4]{The Graduate University for Advanced Studies (SOKENDAI), 1-1 Oho, Tsukuba, Ibaraki 305-0801, Japan}
\affil[5]{Institute of Particle and Nuclear Studies (IPNS), High Energy Accelerator Research Organization (KEK) , 1-1 Oho, Tsukuba, Ibaraki 305-0801, Japan}
\affil[6]{Kavli Institute for the Physics and Mathematics of the Universe, The University of Tokyo, 5-1-5 Kashiwanoha, Kashiwa, Chiba, 277-8583, Japan}
\affil[7]{Laser Zentrum Hannover e.V., Hollerithallee 8, 30419 Hannover, Germany}

\affil[1]{Corresponding author: tmatsumu@astro.isas.jaxa.jp}

\dates{Compiled \today}

\ociscodes{(310.1210) Antireflection coatings; (050.2065) Effective medium 
theory; (050.6624) Subwavelength structures; (050.6875) Three-dimensional fabrication.}

\doi{\url{http://dx.doi.org/10.1364/ao.XX.XXXXXX}}

\begin{abstract}

We report on the first use of laser ablation to make sub-millimeter, broad-band, anti-reflection coatings (ARC)   
based on sub-wavelength structures (SWS) on alumina and sapphire. 
We used a 515~nm laser to produce pyramid-shaped structures with pitch of about 320~\micron\ and total 
height of near 800~\micron. Transmission measurements between 70 and 140~GHz are in agreement with 
simulations using electromagnetic propagation software.  The simulations indicate that SWS ARC with the fabricated
shape should have a fractional bandwidth response of $\Delta \nu / \nu_{center} = 0.55$ centered on 235~GHz for which reflections 
are below 3\%. Extension of the bandwidth to both lower and higher frequencies, between few tens of GHz and few THz, 
should be straightforward with appropriate adjustment of laser ablation parameters.  

\end{abstract}

\setboolean{displaycopyright}{true}

\begin{document}

\maketitle
\thispagestyle{fancy}
\ifthenelse{\boolean{shortarticle}}{\abscontent}{}

\section{Introduction}

The mm and sub-mm regions of the electromagnetic spectrum are important for observations of a variety of astrophysical sources.  
Extinction by galactic dust is low at these wavelengths and therefore otherwise obscured objects,
such as galactic embedded protostar regions, the black hole at the center of the Milky Way, and dusty high redshift galaxies
become observable~\cite{mag_protostar,event_horizon,emission_lines}. Also, the cosmic microwave 
background (CMB) radiation, which encodes a wealth of information about the origin and 
evolution of the Universe~\cite{planck2014-a01}, peaks at mm wavelengths. 
There is a host of instruments conducting observations at these wavelengths.

One of the demands for many astrophysical observations, including those targeting the polarization
of the CMB, is to achieve broad frequency coverage. 
Simultaneous broad bandwidth within a single instrument enables efficient use of telescope and observing time, 
obviating the need to conduct repeated observations each with a different filter bank and/or optical 
elements optimized for the specific frequency band. But 
with broad-bandwidth observations, all optical elements along the light path, including lenses, filters, vacuum windows, 
and polarization modulators (if applicable) should exhibit high optical efficiency, namely, high transmission.
High transmission is achieved by minimizing absorption through appropriate choice of materials and minimizing reflections 
by using an appropriate anti-reflection coatings (ARC).

Sub-wavelength structures (SWS) have been widely explored as ARC in the visible and infrared regions where 
they are often referred to as moth's eye structures (for a review see Raut et al.~\cite{raut2011}). 
SWS-based ARCs have the advantages that they can provide a relatively large operating bandwidth with low reflection
while at the same time (1) there is no need to match indices, thicknesses, and material types between multiple layers of 
different materials, and (2) the structure is robust in cryogenic applications because it is made of the same material as 
the substrate, obviating the need to match thermal coefficients of expansion between different materials. 

Recently SWS ARCs have been implemented in the mm-wave  band
on rexolite \cite{umn_rexolite}, high density polyethylene \cite{almaARC2015}, and silicon \cite{dattaAR}.
In these cases the fabrication approach relied on mechanical machining or injection molding. 
But neither direct mechanical machining  nor injection molding is practical for the small features and high aspect ratios required
when 
the materials are hard or have high melting temperatures, as is the 
case with alumina and sapphire. 

Both alumina and sapphire have optical properties that make them 
appealing for use in mm waveband instruments. They have high index of refraction $n\simeq 3$ giving lenses high 
correcting power with relatively small curvatures and thicknesses; a-cut sapphire's birefringence is near 10\% enabling 
the making of thin retarders; both materials are nearly opaque in the IR and have high thermal conductance making them 
ideal absorbing filters of high frequency radiation in cryogenic instruments. 
(Rosen et al.~\cite{Rosenetal2013} give similar motivations for their development of 
two- and three-layer ARC on alumina based on layered epoxies.)

Laser ablation is an alternative approach for fabricating SWS structures. Several authors reported on 
the combined use of dry chemical etching and laser ablation on silicon to produce ARC
in the visible and near IR~\cite{wu2001,fowlkes2000,her1998,zhao2003,younkin2003}. Here we report on the first 
use of laser ablation to generate millimeter-wave SWS ARC on alumina and sapphire. In our application there 
are no chemical etchants, and the process takes place in regular atmospheric environment. 
In Section~\ref{sec:design} we describe the design of the SWS.  The samples and fabrication process are detailed 
in Section~\ref{sec:sample}.  
Section~\ref{sec:measure} gives our measurements of the geometrical shape and mm-wave transmission of the SWS samples.  
We discuss and summarize our results in Sections~\ref{sec:discussion} and~\ref{sec:summary}. Together with this 
paper, which concentrates on the optical properties of the ablated surfaces, we are referring readers interested
in the technicalities of the laser ablation to a companion paper~\cite{schutz15}.

\section{Design}
\label{sec:design}

\begin{figure}[ht] 
   \centering
   \includegraphics[width=0.8\linewidth]{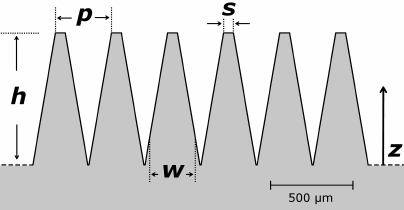} 
   \caption{Cross-sectional view of the designed SWS. They are made by laser ablation 
   of grooves in the substrate. The filling fraction increases toward the substrate, gradually increasing the index.  
   The specific design parameters are: $p=330$~\micron, $h=810$~\micron, and $s=60$~\micron. The width $w$ is a function 
   of the $z$ coordinate. 
   \label{fig:exampleshape} }
\end{figure}

\begin{figure}
   \centering
   \includegraphics[width=0.65\linewidth]{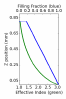} 
   \caption{The fill fraction $f(z)$ (blue solid) and effective index $n_{eff}(z)$ (green dashed) for 
	          the design geometry on alumina 
            with $n_s = 3.1$ (see Figure~\ref{fig:exampleshape} 
            and Table~\ref{tab:geo}). The effective index is calculated using EMT (see text). 
   \label{fig:fz_nz}  }
\end{figure}

Our nominal design for the SWS-ARC is shown in Figure~\ref{fig:exampleshape}. It consists of a 
square array of pyramids each with square cross section. 
As in all SWS-ARCs the goal is to achieve a gradient in the index of refraction through an increase in the area fill fraction of 
the material as radiation propagates in the $-z$ direction.  
With an appropriately designed gradient, reflections can be 
minimized over a broad range of frequencies. The fill fraction for our nominal design, defined
as $f(z) = w(z)/p$, is shown in Figure~\ref{fig:fz_nz}. The approximate
reflection properties of a material with $f(z)$ can be calculated using effective medium theory (EMT)~\cite{brauer94}, 
which gives a conversion from $f(z)$ to an effective index $n_{eff}(z)$; $n_{eff}(z)$ is also shown in Figure~\ref{fig:fz_nz}.
The reflection is then calculated using a standard multi-layer coating approach with a large number of stacked thin layers, each 
with its $n_{eff}$.  

\begin{figure}
   \centering
   \includegraphics[width=\linewidth]{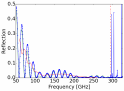}
      \caption{EM-FEA (blue), EMT (green), and EM-FEA averaged across 30\% bandwidth (red) predictions of 
       reflection for the design geometry 
       on alumina, $n_s =3.1$. The total sample thickness, including the SWS, is 4.62 mm, and the 
       SWS coating with the design parameters is assumed on both sides. For EMT we calculated $n_{eff}(z)$ as 
       shown in Figure~\ref{fig:fz_nz} with 150 layers. 
       The vertical red dash line marks the frequency where diffraction is first expected to occur at normal incidence.
   \label{fig:designedshape} }
\end{figure}

Figure~\ref{fig:designedshape} shows the reflection predicted for the 
nominal design using $n_{eff}(z)$ as calculated using EMT and using electromagnetic finite element analysis
(EM-FEA) approach~\cite{hfss}. 
We found that in comparisons to EM-FEA EMT produces reflection values that are 
within 10\% of the numerical calculations as long as the ratio of pitch to wavelength satisfies $p/\lambda \simlt 0.1$.  
Deviations could increase for larger ratios, depending on the specific structures simulated. 
With $p=330$~\micron\ EMT is expected to be accurate (within 10\%) only for frequencies below 
$\sim$90~GHz. Therefore in this paper comparisons to measured data use only EM-FEA. 

At normal incidence the upper band-edge of the SWS-ARC is set by $\nu_u = c / (p n_s) \approx 290$~GHz, above 
which diffraction sets in~\cite{motamedi92}.  This is well reproduced by the EM-FEA calculations as 
sharp reflection spikes; see Figure~\ref{fig:designedshape}. The lower band-edge, defined here as 
the frequency at which band averaged reflections drop below 3\%, is approximately 
$\nu_l \sim c / h \sim 110$~GHz.  We use a bandwidth of $\Delta \nu/\nu = 30\%$ for averaging the reflections. 

\section{Sample Preparation}
\label{sec:sample}

\subsection{Materials}
\label{sec:materials}

We use alumina and c-cut sapphire; the physical parameters of the native
samples are given in Table~\ref{tab:sampleparameters}.
C-cut sapphire is non-birefringent 
for radiation at normal incidence. Both samples are laser ablated on one-side only. 
The index of refraction of the two materials is measured 
using different samples of alumina and sapphire but of the same purity level; they are listed in Table~\ref{tab:sampleparameters}. 

\begin{table*}[t]
   \centering
	 \caption{\bf Measured sample properties and 1$\sigma$ measurement errors.}
   \begin{tabularx}{\textwidth}{p{0.8in}clccc} 
	    \hline
                 & &                    & diameter (mm)  & thickness (mm)      & index\textsuperscript{a}       \\ 
    &  Alumina     & amorphous; 99.6\% pure         & $42$ & $2.21 \pm 0.01$   & $3.04 \pm 0.02 $        \\ 
    &  Sapphire    & c-cut; 99.99 pure               & $51$ & $3.832 \pm 0.003$   & $3.074 \pm 0.003$    \\
		  \hline
   \end{tabularx}
	 \vspace{-.12in}
	 \caption*{\textsuperscript{a}\footnotesize{The indices of refraction values were
               measured on samples with the same purity level that were not laser-ablated. 
               Therefore some difference between the measured indices and the actual indices of our samples is expected. 
               } }
   \label{tab:sampleparameters}
\end{table*}

\subsection{Fabrication}
\label{fabrication}

\begin{figure}
   \centering
   \includegraphics[width=\linewidth]{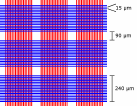} 
   \caption{The ablation pattern consists of $N_{p}$ identical scans across the surface. In each scan the laser makes 
   a single raster pass of the horizontal lines (blue) then the vertical lines (red).  
   The groups of blue/red lines become the grooves between the pyramids. With alumina the lines are 27~mm long and $N_{p} = 200$. 
   With sapphire each line is 3.3~mm long, and the sample is made up of square sub-areas
   machined separately; $N_{p} = 50$. 
   The progression of the ablation on sapphire is shown in Figure~\ref{fig:sem}.     \label{fig:scan} }
\end{figure}

\begin{figure}
   \centering
   \includegraphics[width=\linewidth]{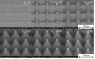} 
   \caption{Scanning electron microscope image of pyramids on a test sapphire sample after $N_{p}$ ablation scans.   
   For the sapphire sample used for the transmission measurements (Fig.~\ref{fig:resultsS3}) we used $N_{p}=50$ 
   because there was only minor evolution in structure shape for subsequent passes. \label{fig:sem}  }
\end{figure}

Ablation is done using a 515 nm laser operating with 7 ps pulses and a repetition rate of 400~kHz. 
At focus the beam has a $1/e^{2}$ width of 
30~\micron. The beam is scanned in a raster pattern across the surface, first in one direction then the orthogonal, as shown 
schematically in Figure~\ref{fig:scan}. The progression of the ablation as a function of the number
of passes is shown in Figure~\ref{fig:sem}. 
The alumina was machined using 200 identical passes, covering the center $27 \times 27$ mm$^{2}$ of the sample.  
Machining time was 5 hours.  
The full surface of sapphire was machined in a series of $3.3 \times 3.3$ mm$^{2}$ sub-areas.  
Each sub-area was scanned 50 times using the same scan pattern as in Figure~\ref{fig:scan} before 
moving to the next sub-area.  Total machining time was 5.6 hours. 
Sch{\"u}tz et al.~\cite{schutz15} describe the machining process in more detail. 

\section{Measurements}
\label{sec:measure}

\subsection{Shape}

\begin{figure}
   \centering
   \includegraphics[width=\linewidth]{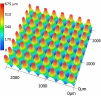} 
   \caption{Optical image of SWS-ARC on sapphire reconstructed from a series of 2-dimensional images taken along the $z$ axis.   
    \label{fig:S3_keyence}  }
\end{figure}

\begin{figure}
   \centering
   \includegraphics[width=\linewidth]{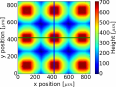} 
   \caption{Height map of SWS on sapphire. The 1-dimensional height profiles that are shown 
   in Figure~\ref{fig:S3_cuts} are marked here as broad blue and green 
   lines.  The width of each line corresponds to the portion of the height map 
   averaged to produce the 1-dimensional profile. 
   \label{fig:S3_height}  }
\end{figure}

We use a Keyence VHX-5000 optical microscope  and a Nikon A1RMP confocal microscope  to 
image the laser ablated surfaces. The microscopes take a series of 2-dimensional images spaced equally in~$z$. The 
in-plane resolution is 1.7~$\mu$m or higher and depends on the specific microscope.  
The spacing in $z$ is 10~$\mu$m for the optical microscope and $1.3-4$~$\mu$m 
for the confocal microscope, depending on image location. This series of images is used to 
reconstruct a 3-dimensional image of the sample. A section of the sapphire sample is shown in perspective 
in Figure~\ref{fig:S3_keyence}. A zoomed version shown in projection along the $z$ axis is in Figure~\ref{fig:S3_height}. 

Using the optical images we measured the parameters of the machined surfaces in ten different locations for 
each of the alumina and sapphire samples. In each location we image a square consisting of 
approximately two pyramids on a side and extract the geometric parameters of the machined samples. The average
values across all these spatial locations are given in Table~\ref{tab:geo}.  
Because groove depth is not uniform - grooves are 
deepest at groove crossings - there may be ambiguity about the height measurement.
To measure height we first calculate the fill fraction as a function of $z$ in steps of 5 \micron\ and set $z=0$ 
when the fill fraction is 1. maximum height is defined by the $z$ position where the fill fraction is 0.  
For all values in Table~\ref{tab:geo}, the errors quoted are the standard deviation in the measurements 
across the different areas and provide a measure of structure uniformity. 

\begin{figure}
   \centering
   \includegraphics[width=0.8\linewidth]{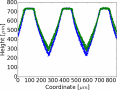} 
   \caption{1-dimensional height profiles on sapphire. The profiles are produced from cuts through peak centers, as shown 
            in Figure~\ref{fig:S3_height},
            for 6 locations across the sample. Cuts at constant $y$ are in green; constant $x$ are in blue. 
            They have been manually translated to align vertically and horizontally to enable visual comparison of structure uniformity. 
            \label{fig:S3_cuts}	 }
\end{figure}

\begin{figure}
   \centering
   \includegraphics[width=\linewidth]{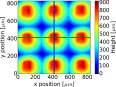} 
   \caption{Height map of SWS on alumina. The 1-dimensional height profiles that are shown 
   in Figure~\ref{fig:A1_cuts} are marked here as broad blue and green 
   lines.  The width of each line corresponds to the portion of the height map 
   averaged to produce the 1-dimensional profile. 
    \label{fig:A1_height}  }
\end{figure}

\begin{figure}
   \centering
   \includegraphics[width=0.8\linewidth]{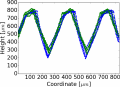} 
   \caption{1-dimensional height profiles on alumina. The profiles are produced from cuts 
	          through peak centers, as shown in Figure~\ref{fig:A1_height},
            for 5 locations across the sample.  
            Cuts at constant $y$ are in green; constant $x$ are in blue. 
            They have been manually translated to align vertically and horizontally to enable 
						visual comparison of structure uniformity.             
    \label{fig:A1_cuts}	 }
\end{figure}

We characterize the symmetry of the structures by comparing 1-dimensional height profiles in two orthogonal directions 
that are parallel to the grooves. These 1-dimensional profiles on sapphire (alumina) are illustrated in Figure~\ref{fig:S3_height} 
(\ref{fig:A1_height}) and are plotted in Figure~\ref{fig:S3_cuts} (\ref{fig:A1_cuts}).  
We discuss these measurements in Section~\ref{sec:discussion}.

\begin{table}
   \centering
	 \caption{\bf Geometric parameters of SWS. }
   \begin{tabularx}{\linewidth}{cccc}
	  \hline
	  Sample           & height ($\mu$m) & pitch ($\mu$m) & peak width  ($\mu$m) \\
		\hline
		   \multicolumn{4}{c}{Designed} \\
		All          & $810 $ &  $330$   & $60$   \\
		\hline
		   \multicolumn{4}{c}{Measured\textsuperscript{a}} \\
		Alumina          & $790\pm60$  &  $313 \pm 4$   & $66 \pm 8$   \\
		Sapphire         & $715\pm 24$ &  $325 \pm 4$   & $57 \pm 6$   \\
		\hline
   \end{tabularx}
    \vspace{-0.12in}
   	 \caption*{\textsuperscript{a}\footnotesize{Error quoted is the standard deviation for measurements in 10 locations. Individual
	 measurement errors are $\pm$2~$\mu$m for pitch and $\pm$4~$\mu$m for height and peak width. }}
	 \label{tab:geo}
\end{table}

\subsection{Transmission}

\begin{figure*}[ht] 
   \centering
   \includegraphics[width=0.82\linewidth]{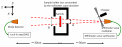} 
   \caption{Diagram of the experimental setup.  Aperture diameters for sapphire were $a=3$~cm and $b > 10$~cm.  
	  For alumina $a=1.5$~cm and $b > 10$~cm.
   \label{fig:experimentalsetup} }
\end{figure*}

Transmission is measured at room temperature using a millimeter wave source and a power detector. 
A schematic diagram of the setup is 
shown in Figure~\ref{fig:experimentalsetup}. 
Two adjustable-frequency mm-wave multipliers are used to up-convert the GHz signal of the generator 
to bands between 75 and 110 GHz, and 90 to 140 GHz. Pyramidal horns with 3~dB beam divergence of 
$\sim$25~degrees couple the linearly polarized mm-wave 
signal to free space. The source is chopped with a mechanical blade producing a 
modulated signal at 10~Hz. We use a lock-in amplifier at the output of the detector.  
The sample is contained within a box that has two apertures and is lined with mm-wave absorbing material. 
The polarization axis of the mm-wave source is aligned parallel to one of the principal directions of 
the SWS-ARC, as defined by the orientation of the grooves.  The sample can be rotated by 90 degrees to 
enable transmission measurements at two orthogonal polarizations.

\begin{figure}
   \centering
   \includegraphics[width=\linewidth]{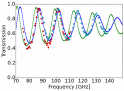} 
   \caption{Measured transmission as a function of frequency for sapphire at two frequency 
   bands between 75 and 110~GHz (red) and between 90 and 140~GHz (cyan). All data were taken 
   at a single polarization state. We show HFSS predictions for a sample that is ablated on one side and 
   assuming all SWS are made with the tallest ($h=740$~\micron , blue) 
   and shortest ($h=680$~\micron , green) profiles measured 
   in 10 different locations across the sample. 
   The data are normalized to the blue HFSS curve at 87~GHz. 
	 \label{fig:resultsS3} }

\end{figure}

\begin{figure}
   \centering
   \includegraphics[width=\linewidth]{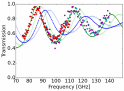} 
   \caption{Measured transmission as a function of frequency for alumina at two frequency 
   bands between 75 and 110~GHz (red, a single polarization state) 
   and between 90 and 140~GHz (cyan for one polarization state, and purple for the orthogonal polarization 
   state). We show HFSS predictions for a sample that is ablated on one side and 
   assuming all SWS are made with the tallest ($h=875$ \micron , blue) 
   and shortest ($h=715$ \micron ,green) profiles measured in 10 different locations across the sample, and for  
   $n_{s}=3.16$ (dark blue and green) and $n_{s}=3.04$ (light blue and green). 
   The data are normalized to the dark green HFSS simulation at 85~GHz. 
   \label{fig:resultsA1} }
\end{figure}

A measurement consists of a relative normalization stage in which the sample 
is not present, but the sample mount and all apertures are included, and a data taking stage with the sample. 
During the normalization stage the output of the power detector is logged as a function of the source frequency $\nu$
and detector position along the $z$-axis. The range of sampling in $z$ is slightly larger than one wavelength. These are 
the normalization data $N(\nu)$. The sample is then 
inserted into its holder and data are retaken at the same source frequencies and $z$ positions.
These are the data $D(\nu)$. For each frequency $\nu$ both $N$ and $D$ are fit to models 
\begin{equation}
N(\nu,z) = N_{0}(\nu) + N_{1}(\nu) \sin{ (N_{2}(\nu) z + N_{3}(\nu)), }
\end{equation}
 and 
 \begin{equation}
 D(\nu,z) = D_{0}(\nu) + D_{1}(\nu) \sin{ (D_{2}(\nu) z + D_{3}(\nu)) }, 
 \end{equation}
where $N_{i},\,\, D_{i}$ ($i=0,1,2,3$) are the fit parameters. 
The transmission data we report is 
\begin{equation}
T(\nu) = C \frac{D_{0}(\nu)}{N_{0}(\nu)}, 
\end{equation}
where $C$ is a constant used to normalize $T$ to the EM-FEA predictions, as discussed in Section~\ref{sec:discussion}. 
The z-dependent sinusoidal fits account for standing waves between the source and detector. 
The transmission data for the two samples is shown in Figures~\ref{fig:resultsS3} and~\ref{fig:resultsA1} along with 
HFSS predictions, which are discussed in Section~\ref{sec:discussion}. Generally, data were taken at 
one polarization state. With alumina data at the higher frequency band was recorded at two polarization states.  

\section{Discussion}
\label{sec:discussion}

\subsection{Geometry}
\label{sec:discussion_geometry}

The spatially averaged shape parameters of both samples match the nominal design to better than 12\%; 
the height of the sapphire pyramids is consistently short by 12\% and the width of the alumina tips are too wide
by 11\%, although there is a similar level of spatial dispersion across the sample, 
see Table~\ref{tab:geo}. Otherwise the shape parameters agree within 5\%. 
This level of agreement with the design was achieved with a small number of 
trials. No extensive attempts were made to decrease discrepancies between the designed and ablated 
structures. With single crystals like sapphire better agreement is achievable and we have not identified any limitation that would 
prevent even more accurate ablation. Alumina is sintered material and its properties may show variations between 
different samples. This aspect requires further investigation. Variation in laser power is not a likely 
cause for the $\sim$10\% spatial dispersion of the measured
parameters as power was stable to better than 0.1\% over the 5-6 hours it took to ablate each sample.

The ablated SWS have square tops and they transition to approximately circular bases. The structures 
are slightly asymmetric; saddles between peaks in one orientation are on average slightly deeper than 
in the orthogonal direction, see Figures~\ref{fig:S3_cuts} and~\ref{fig:A1_cuts}. Such asymmetry, while 
avoidable with better tuning of ablation parameters, may be important for polarimetric studies. 

We quantify the symmetry of the pyramids by calculating the full width at half height (FWHH) 
and the saddle depth $sd$ for $x$ and $y$ slices at each of the ten image locations on the samples.  
We find the global maximum of each image, excluding the highest 1\% of points to 
avoid outliers, and the depth of each saddle for each slice ($sd$), then determine the width of 
each pyramid at the halfway point between the maximum and the saddle (FWHH). 
The procedure produces values for FWHH$_x$, $sd_x$, FWHH$_y$, and $sd_y$.
We form a ratio 
\begin{equation}
r_{sym} = \frac{\text{FWHH}_x}{\text{FWHH}_y},
\end{equation}
so symmetric structures have $r_{sym} = 1$.
Table \ref{tab:fwhh} gives average values and standard deviations for $r_{sym}$ and $sd$.  
Some structure asymmetry is expected when identical ablation 
parameters are used for the two orthogonal directions. Relative to the first pass, the pass in the orthogonal 
direction will produce on average deeper grooves. This 
asymmetry can be mitigated by tuning the scan speed or number of scan repeats. 
\begin{table}[h]
   \centering
    \caption{\bf Average symmetry parameter $r_{sym}$ and saddle depth $sd_{x,y}$ for the two samples.  
    The $1 \sigma$ standard deviations reflect spatial variations across the sample.  
     }
   \begin{tabular}{cccc}
	  \hline
	             & $r_{sym}$         & sd$_x$ ($\mu$m)   & sd$_y$ ($\mu$m) \\
   	  Alumina    & $1.042 \pm 0.035$ &  $ 545 \pm 50 $  & $ 525 \pm 25 $ \\
	  Sapphire   & $0.950 \pm 0.015$ &  $ 500 \pm 10 $  & $ 450 \pm 10 $ \\
		\hline
   \end{tabular}
	 \label{tab:fwhh}
\end{table}

\subsection{Transmission}

The measured transmission for each sample is plotted in Figures \ref{fig:resultsS3} and \ref{fig:resultsA1} along 
with EM-FEA simulations.
The simulated transmission is calculated by importing the measured height map of a single peak into HFSS,
mirroring it across the $x$-axis, and then mirroring the resulting pair across the $y$-axis to create an array 
of four peaks whose opposing boundaries are identical.
This allows us to implement periodic boundary conditions and simulate the sample as an 
infinite plane.\footnote{The mirroring operation produces a repetition of structures on a length scales of $2p$.  
This artificial periodicity creates diffraction features at a frequency $\nu = 2c/ (p n_{s}) $, which 
we manually remove from the Figures.} 
For each sample we indicate a possible range of predictions by simulating the transmission for 
a sample made with the tallest and shortest structures found by imaging 10 locations 
on the samples. The phase and amplitude of the transmission interference fringes vary between the two cases,
with the phase difference showing larger sensitivity to height difference. The frequency of the fringe pattern
is sensitive to the total thickness of the material. With taller structures, 
the non AR-coated part of the sample is thinner, resulting in a lower frequency fringe pattern. 

With sapphire the data and measured shape variability match EM-FEA predictions closely. The data is bracketed 
by prediction with the tallest and shortest structures measured. This is not the case with alumina for which the data
is outside the range of shortest to tallest predictions made 
with the measured index of $n_{s} = 3.04$; see light green to blue in Figure~\ref{fig:resultsA1}. 
We note that reported measurements of the index of alumina have
larger variability than those with sapphire~\cite{lamb96} and that our measurement was carried out 
on a different sample than the one ablated. The index of alumina that better matches our data is $n_{s} = 3.16$
and this value is within the range of variability reported in the literature~\cite{lamb96}.  
Predictions with this index for the shortest and tallest structures measured 
are shown as bold in Figure~\ref{fig:resultsA1}. 

\begin{figure}[htbp] 
   \centering
   \includegraphics[width=\linewidth]{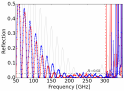} 
   \caption{Simulated reflection of our machined ARC applied to both sides of a sapphire flat.  The blue curve 
	          corresponds to the structures and polarization plotted in blue in Figure~\ref{fig:resultsS3}.  
						The orthogonal polarization for the same structures is in red.  The vertical red line 
	          indicates where $p = \lambda/n_s$ and the structures can no longer be considered sub-wavelength.
						A basic $\lambda /4$ ARC optimized at 250~GHz is shown in gray.}
   \label{fig:S3_300GHz}
\end{figure}

\begin{figure}[htbp] 
   \centering
   \includegraphics[width=\linewidth]{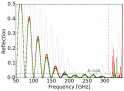} 
   \caption{Simulated reflection of our machined ARC applied to both sides of an alumina flat.  The green curve 
	          corresponds to the structures and polarization plotted in green in Figure~\ref{fig:resultsA1}. 
						The orthogonal polarization for these is in red. Vertical red line indicates the point where 
						$p = \lambda/n_s$ and the structures can no longer be considered sub-wavelength.}
   \label{fig:A1_300GHz}
\end{figure}

We can asses the expected broadband performance of the fabricated SWS by extending 
the HFSS simulations to 350~GHz
and assuming that the sapphire and alumina samples are coated on {\em both} sides. The simulations
assume the {\it measured} sapphire (alumina) pyramid shape with a height of 740~\micron\ (715~\micron)
and follow the same procedure described earlier to create periodic boundary conditions. 
The predictions are shown in Figures~\ref{fig:S3_300GHz} and~\ref{fig:A1_300GHz}, each for two incident polarization states.
Starting at low frequencies, reflection decreases with increasing frequency, dropping below~3\% at 170~GHz for both 
polarizations and both materials. On Sapphire, 
a total bandwidth of 130~GHz is achieved with band averaged reflection of 1.0\%. 
The high frequency 
side is limited by sharp reflection features that are introduced by diffraction. We discuss these below. 
There is a slightly larger effective bandwidth of 140~GHz on alumina because it has a smaller pitch, moving 
the sharp reflection features to higher frequencies.  
The band averaged reflection of alumina between 170 and 310~GHz, is also 1.0\%.
For comparison, a standard $\lambda/4$ coating optimized for 250~GHz, also shown in the Figures (light gray), 
has a bandwidth of only 60~GHz. 
Rosen et al.~\cite{Rosenetal2013} report bandwidths $\Delta \nu / \nu$ of 92\% and 104\% with reflection below 
10\% with two- and three-layer ARC on alumina using layered epoxies. Their 
center frequency is $\sim$150~GHz. With this first round of laser ablation, our alumina sample
has  $\Delta \nu / \nu= 74\%$, centered on 225~GHz, with reflections below 10\%.

At frequencies between 100 and 170~GHz the reflection expected from the fabricated structures 
is higher than predictions based on the design geometry; 
compare to Figure~\ref{fig:designedshape}. This is because the 
design geometry has a uniform pyramid height of 810 \micron; in other words the grooves have 
uniform depth of 810 \micron\ relative to the tips. Yet with the ablated structures the maximum 
groove depth is achieved over a smaller fraction of groove length. The {\it average} groove depth relative to the 
peaks is $\sim$550~\micron (600~\micron) on sapphire (alumina),  
which is a factor of 1.5 (1.4) shorter than the original design.
Recalling that the lower band cut-off frequency is proportional to $1/h$ we should find 
that the lower frequency edge of the band is a factor of $\sim$1.5 higher than the anticipated 110~GHz, in good agreement
with the EM-FEM calculations. 

The bandwidth of the SWS-ARC is extendable to lower frequencies by increasing the height of the structures 
and to higher frequencies by decreasing the pitch. A useful figure of merit is the aspect ratio $a = h/p$. The current
value of $a$, using $h_{eff} = 550$~\micron\ and $p=315$~\micron\ is 1.7.  Increasing the depth to 
$h_{eff}=1100$~\micron, thus doubling the aspect ratio, would halve the low frequency edge of the band giving 
a useful band between $\sim$85 and 300~GHz. The same aspect ratio but produced with half the pitch would 
give a band between 170 and 600~GHz. We see no physical limitation for increasing the aspect ratio by a factor of 5-10. 

The simulations shown in Figures ~\ref{fig:S3_300GHz} and~\ref{fig:A1_300GHz} reflect the 
height and symmetry differences between the samples. The somewhat higher asymmetry with the 
sapphire sample gives rise to a somewhat stronger differential reflection, as noted primarily by the phase 
difference in the reflection fringes between the two polarization states. However, averaging over a 
$\Delta \nu / \nu = 0.3$ band that is centered on 235~GHz, as would be expected with a broadband instrument
observing in that band, we find a differential reflection of less than 0.1\%
for both materials. 

Theory and HFSS simulations indicate that at normal incidence 
diffraction at wavelengths shorter than $\lambda_{u} \sim p n_{s}$ produces
strong reflection features. The density of these features increases with increasing frequency. 
At off-normal incidence angles the onset of the reflection features shifts to lower frequencies. 
To avoid such reflections, experiments using SWS-ARC should implement low pass filters to block
high frequency radiation. 

\section{Summary}
\label{sec:summary}

We have demonstrated the first use of laser ablation to produce millimeter-wave SWS ARC on hard, 
difficult-to-machine alumina and sapphire.  The ablation is carried out in regular atmospheric conditions 
and involves no additional chemical etchants. Structures with aspect ratio of 1.7, average height of 
$\sim$550~\micron, and maximum height of $\sim$800~\micron\ were fabricated on one 
side of $\sim$5 cm diameter samples within about 5 hours. Measurements
of transmission are in close agreement with electromagnetic simulations. The simulations indicate that 
with SWS ARC on two sides these samples have reflections of less than 3\% over a band between 170 and 300~GHz, a 
fractional bandwidth of $\Delta \nu / \nu_{center} =0.55$. Significantly larger bandwidths and faster ablation 
rates are possible by appropriately tuning laser parameters. A companion paper by Sch\"utz et. al~\cite{schutz15} 
gives more information about the laser ablation process.

\section*{Acknowledgement}
The authors acknowledge use of resources provided by the
Minnesota Nanofabrication Center (\url{http://nfc.umn.edu}), the University of Minnesota Imaging Center (\url{http://uic.umn.edu}), and 
the Minnesota Supercomputing Institute (\url{http://www.msi.umn.edu}). TM thanks Prof. M. Hasegawa for use of the 
millimeter-wave source. This work was partially supported by JSPS KAKENHI grant numbers 24740182 
and 15H05441, the Mitsubishi foundation (grant number 24 in JFY2015 in Science and technology) and by the ISAS strategic development fund of the steering committee for space science.

\bibliography{laserARC}

\end{document}